\documentclass{article}
\usepackage[T1]{fontenc}

\makeatletter

\newcommand{\LyX}{L\kern-.1667em\lower.25em\hbox{Y}\kern-.125emX\spacefactor1000}

\makeatother

\begin{document}

\begin{titlepage} 

\begin{flushright} {\small DFTT-47/98} \\ hep-th/9808146 \end{flushright} 

\vfill 

\begin{center} {

{\large All roads lead to Rome:\\

Supersolvables and Supercosets} 

\vskip 0.3cm

{\large {\sl }} 

\vskip 10.mm 

{\bf I. Pesando }\footnote{e-mail: ipesando@to.infn.it, pesando@alf.nbi.dk\\

Work supported by the European Commission TMR programme ERBFMRX-CT96-004}

\vskip4mm

{\small Dipartimento di Fisica Teorica , Universi\'a di Torino, via P. Giuria 1, I-10125 Torino, Istituto Nazionale di Fisica Nucleare (INFN) - sezione di Torino, Italy} 

}

\end{center} 

\vfill

\begin{center}{ \bf ABSTRACT}\end{center} 

\begin{quote}

We show explicitly that the two recently proposed actions for the type IIB superstring propagating on \( AdS_{5}\times S_{5} \) agree completely. 


\vfill

\end{quote} 

\end{titlepage}

\newcommand{\hepth}[1]{hep-th/#1}

In the last period there has been quite a lot of activity in finding the GS
action for a type IIB string propagating on \( AdS_{5}\times S_{5} \), this
has resulted in two proposals which are apparently different.

It is purpose of this letter to show that they are exactly the same thus clarifying
some doubts.

The first proposed action (\cite{IP}) reads

\begin{eqnarray}
 & S=\int d^{2}\xi \, \sqrt{-g}\, g^{\alpha \beta }\frac{1}{2}\times  & \nonumber \\
 & \left\{ \eta _{pq}\frac{\rho ^{2}}{e^{2}}\left[ \partial _{\alpha }x^{p}+\frac{ie}{4}\left( \vartheta ^{\dagger N}\widetilde{\sigma }^{p}\partial _{\alpha }\theta _{N}-\partial _{\alpha }\vartheta ^{\dagger N}\widetilde{\sigma }^{p}\theta _{N}\right) \right] \left[ \partial _{\beta }x^{q}+\frac{ie}{4}\left( \vartheta ^{\dagger M}\widetilde{\sigma }^{q}\partial _{\beta }\theta _{M}-\partial _{\beta }\vartheta ^{\dagger M}\widetilde{\sigma }^{q}\theta _{M}\right) \right] \right. \,  & \nonumber \\
 & \left. -\frac{1}{e^{2}}\frac{\partial _{\alpha }\rho \partial _{\beta }\rho }{\rho ^{2}}-\frac{4}{e^{2}}\delta _{ij}\frac{\partial _{\alpha }z^{i}\partial _{\beta }z^{j}}{(1-z^{2})^{2}}\right\}  & \nonumber \\
 & -\frac{i}{4e}\rho \: \left[ d\theta ^{\dagger N}\sigma _{2}d\theta ^{*M}\, \overline{\eta }_{N}(z)\eta _{cM}(z)+d\theta ^{T}_{N}\sigma _{2}d\theta _{M}\, \overline{\eta }_{c}^{N}(z)\eta ^{M}(z)\right]  & \label{mia} 
\end{eqnarray}
and it was obtained using the supersolvable algebra tecnique (\cite{gruppoTo})
while the other (\cite{Kall-act}\cite{Kall-Tsey}\cite{MatsaevTseytlin}\cite{Kallosh}\cite{Kall-fixing})
reads

\begin{eqnarray}
S=-\frac{1}{2}\int d^{2}\xi \,  & \left[ \sqrt{-g}\, g^{\alpha \beta }y^{2}\left( \partial _{\alpha }x^{p}+2i\: \widehat{\overline{\vartheta }}\Gamma ^{p}\partial _{\alpha }\widehat{\theta }\right) \left( \partial _{\beta }x_{p}+2i\: \widehat{\overline{\vartheta }}\Gamma _{p}\partial _{\beta }\widehat{\theta }\right) +\frac{1}{y^{2}}\partial _{\alpha }y^{t}\partial _{\beta }y_{t}\right.  & \nonumber \\
 & \left. +4i\epsilon ^{\alpha \beta }\: \partial _{\alpha }y^{t}\: \widehat{\overline{\vartheta }}\Gamma _{t}\partial _{\beta }\widehat{\theta }\right]  & \label{loro} 
\end{eqnarray}
 Let us now specify the notations and spell out the differences of the two actions.
The first thing to notice and the most trivial is that the first one uses a
mostly minus metric while the second one uses a mostly plus metric. Second the
indeces run as follows \( p,q,\ldots =0,\ldots ,3 \), \( i,j,\ldots =5,\ldots ,9 \)
, \( t,u,\ldots =4,\ldots ,9 \) because the first one uses horospherical coordinates
\( \left\{ x^{p},\rho \right\}  \) on \( AdS_{5} \) and projective \( \left\{ z^{i}\right\}  \)
coordinates on \( S_{5} \) while the second one uses cartesian coordinates
\( \left\{ x^{p},y^{t}\right\}  \). But the most striking difference is in
the fermionic sector: the fermionic coordinates in ( \ref{mia}) \( \theta _{N} \)
are a set of \( N=4 \) Weyl spinor in \( D=4 \) (or that is the same half
a spinor in \( D=5 \) : this is the effect of fixing the \( \kappa  \) symmetry
on \( AdS_{5|4} \) ) while the ones used in (\ref{loro}) \( \widehat{\theta } \)
are a Majorana -Weyl spinor in \( D=10 \) . In addition to this in (\ref{mia})
enter the c-number Killing spinors on \( S_{5} \) \( \eta ^{N} \) and their
conjugate \( \eta ^{N}_{c}\equiv C_{5}\eta ^{\dagger }_{N} \) \footnote{
We use the following 10D \( \Gamma  \) representation in terms of the \( AdS_{5} \)
\( \gamma ^{a} \) matrices and of the corresponding \( S_{5} \) \( \tau ^{i} \)
matrices 
\[
\Gamma \widehat{^{a}}=\left\{ \gamma ^{a}\otimes 1_{4}\otimes \sigma _{1}\, ,\, 1_{4}\otimes \tau ^{i}\otimes (-\sigma _{2})\right\} \]

with \( \widehat{a},\widehat{b},\ldots =0\ldots 9 \) , \( a,b,\ldots =0\ldots 4 \)
and \( i,j,\ldots =5\ldots 9 \).

And we write the 10D charge conjugation as

\[
\widehat{C}=C\otimes C_{5}\otimes \sigma _{2}\]

where \( C \) , \( C_{5} \) are the \( AdS_{5} \) and \( S_{5} \) charge
conjugation matrices.

All \( C \) .s , \( C^{-1}\gamma ^{a} \) , \( C^{-1}_{5}\tau ^{i} \) are
antisymmetric while \( \widehat{C}^{-1}\Gamma ^{\widehat{a}} \) are symmetric.

Moreover we use the following (1+4)D \( \gamma  \) explicit representation

\[
\gamma _{p}=\left( \begin{array}{cc}
 & \sigma _{p}\\
\widetilde{\sigma _{p}} & 
\end{array}\right) \; \gamma _{4}=\left( \begin{array}{cc}
i\, 1_{2} & \\
 & -i\, 1_{2}
\end{array}\right) \; C=\left( \begin{array}{cc}
i\sigma _{2} & \\
 & i\sigma _{2}
\end{array}\right) \]

with \( p,q,\ldots =0\ldots 3 \), \( \sigma _{p}=\left\{ 1_{2},-\sigma _{1},-\sigma _{2},-\sigma _{3}\right\}  \)
and \( \widetilde{\sigma }_{p}=\left\{ 1_{2},\sigma _{1},\sigma _{2},\sigma _{3}\right\}  \)

and the following 5D \( \tau  \)

\[
\tau _{5}=i\gamma _{0}\; \tau _{6,\ldots ,9}=\gamma _{1,\ldots ,4}\; C_{5}=C\]
 
}which satisfy the equation
\[
D_{SO(5)}\eta ^{N}\equiv \left( d-\frac{1}{4}\varpi ^{ij}\tau _{ij}\right) \eta ^{N}=-\frac{e}{2}\tau _{i}\eta ^{N}E^{i}\]

In order to compare the two actions we need the explicit form of these Killing
spinors; this is not too hard to obtain with the help of the ansatz

\[
\eta ^{N}=\left( a(z^{2})+b(z^{2})\: z^{i}\tau _{i}\right) \epsilon ^{N}\]
 where \( \epsilon ^{N} \) are constant spinors. The result is\footnote{
Our conventions for the \( S_{5} \) coset manifold with the Killing induced
metric, i.e. negative definite, are

\begin{eqnarray*}
dE^{i}-\varpi ^{i}_{.j}E^{j} & = & 0\\
d\varpi ^{ij}-\varpi ^{ik}\varpi _{k}^{.j} & = & e^{2}E^{i}E^{j}
\end{eqnarray*}
 where all the potentials depend on the coordinate \( z \). Explicitly we have
( \( z^{2}=\eta _{ij}z^{i}z^{j}=-\delta _{ij}z^{i}z^{j} \) )

\begin{eqnarray*}
E^{i} & = & \frac{2}{e}\frac{dz^{i}}{1-z^{2}}\\
\varpi ^{ij} & = & \frac{4z^{[i}dz^{j]}}{1-z^{2}}
\end{eqnarray*}

}

\begin{eqnarray}
\eta ^{N} & = & \frac{1}{\sqrt{1-z^{2}}}\left( 1-z^{i}\tau _{i}\right) \epsilon ^{N}\nonumber \\
\eta ^{N}_{c} & = & \frac{1}{\sqrt{1-z^{2}}}\left( 1+z^{i}\tau _{i}\right) \epsilon ^{N}\label{killingS5} 
\end{eqnarray}
with \( \epsilon ^{\dagger }_{N}\epsilon ^{M}=\delta ^{M}_{N} \) in order to
satisfy the normalisation condition \( \eta ^{\dagger }_{M}\eta ^{N}=\delta ^{N}_{M} \);
we can therefore choose the normalisation \( \epsilon ^{N}_{\alpha }=\delta ^{N}_{\alpha } \)
(\( \alpha  \) is the \( 4D \) spinor index). 

When we insert this expression in (\ref{mia}) we get

\begin{eqnarray}
 & S=\int d^{2}\xi \, \sqrt{-g}\, g^{\alpha \beta }\frac{1}{2}\times  & \nonumber \\
 & \left\{ \eta _{pq}\frac{\rho ^{2}}{e^{2}}\left[ \partial _{\alpha }x^{p}+\frac{ie}{4}\left( \vartheta ^{\dagger N}\widetilde{\sigma }^{p}\partial _{\alpha }\theta _{N}-\partial _{\alpha }\vartheta ^{\dagger N}\widetilde{\sigma }^{p}\theta _{N}\right) \right] \left[ \partial _{\beta }x^{q}+\frac{ie}{4}\left( \vartheta ^{\dagger M}\widetilde{\sigma }^{q}\partial _{\beta }\theta _{M}-\partial _{\beta }\vartheta ^{\dagger M}\widetilde{\sigma }^{q}\theta _{M}\right) \right] \right. \,  & \nonumber \\
 & \left. -\frac{1}{e^{2}}\frac{\partial _{\alpha }\rho \partial _{\beta }\rho }{\rho ^{2}}-\frac{4}{e^{2}}\delta _{ij}\frac{\partial _{\alpha }z^{i}\partial _{\beta }z^{j}}{(1-z^{2})^{2}}\right\}  & \nonumber \\
 & -\frac{i}{4e}\rho \: \frac{1+z^{2}}{1-z^{2}}\left[ d\theta ^{\dagger N}\sigma _{2}d\theta ^{*M}\, \epsilon ^{\dagger }_{N}C_{5}\epsilon _{M}^{*}+d\theta ^{T}_{N}\sigma _{2}d\theta _{M}\, \epsilon ^{T\, N}C^{-1}_{5}\epsilon ^{M}\right]  & \nonumber \\
 & -\frac{i}{4e}\rho \: \frac{2z^{i}}{1-z^{2}}\left[ d\theta ^{\dagger N}\sigma _{2}d\theta ^{*M}\, \epsilon ^{\dagger }_{N}\tau _{i}C_{5}\epsilon _{M}^{*}-d\theta ^{T}_{N}\sigma _{2}d\theta _{M}\, \epsilon ^{T\, N}C^{-1}_{5}\tau _{i}\epsilon ^{M}\right]  & \nonumber \label{mia1} 
\end{eqnarray}
The comparison between the two bosonic kinetic terms and the two WZ terms suggests
the following change of variables
\begin{eqnarray}
y^{4} & = & \rho \frac{1+z^{2}}{1-z^{2}}\nonumber \\
y^{i} & = & \rho \: \frac{2z^{i}}{1-z^{2}}\label{bos-chan} 
\end{eqnarray}
In this way the bosonic part of the two actions (\ref{mia}) and (\ref{loro})
agrees perfectly; in particular we get:
\[
\frac{dy^{2}}{y^{2}}=\frac{{d\rho ^{2}}}{\rho ^{2}}-4\: \frac{dz^{2}}{\left( 1-z^{2}\right) ^{2}}\]

We are left with the task of making a \( 10D \) Majorana-Weyl spinor \( \widehat{\theta } \)
out of 4 \( 4D \) Weyl spinors \( \theta _{N} \) and 4 constant \( 5D \)
spinors \( \epsilon ^{N} \) . To this purpose we notice that with our conventions
the following \( 10D \) spinor is Majorana-Weyl for all the \( 4D \) spinors
\( \alpha _{N} \) :

\[
\Theta _{MW}=\left( \begin{array}{c}
\left( \begin{array}{c}
\alpha _{N}\\
\sigma _{2}\alpha ^{*M}\, C_{5NM}
\end{array}\right) \otimes \epsilon ^{N}\\
0_{16}
\end{array}\right) \]

We can now compute all the relevant two fermions currents \( \overline{\Theta }_{MW\, 1}\Gamma ^{\widehat{a}}\Theta _{MW\, 2} \)
using the explicit expression for \( \epsilon _{N} \) explictly:

\begin{eqnarray*}
\overline{\Theta }_{MW\, 1}\Gamma ^{p}\Theta _{MW\, 2} & = & \alpha _{1}^{\dagger N}\widetilde{\sigma }^{p}\alpha _{N\, 2}-\alpha ^{\dagger N}_{2}\widetilde{\sigma }^{p}\alpha _{N\, 1}\\
\overline{\Theta }_{MW\, 1}\Gamma ^{4}\Theta _{MW\, 2} & = & i\: \alpha _{1}^{\dagger N}\sigma _{2}\alpha _{2}^{*M}\; \epsilon ^{\dagger }_{N}C_{5}\epsilon ^{*}_{M}-i\: \alpha ^{T}_{N\, 1}\sigma _{2}\alpha _{M\, 2}\; \epsilon ^{T\, N}C^{-1}_{5}\epsilon ^{M}\\
\overline{\Theta }_{MW\, 1}\Gamma ^{i}\Theta _{MW\, 2} & = & -i\: \alpha _{1}^{\dagger N}\sigma _{2}\alpha _{2}^{*M}\; \epsilon ^{\dagger }_{N}\tau ^{i}C_{5}\epsilon ^{*}_{M}-i\: \alpha ^{T}_{N\, 1}\sigma _{2}\alpha _{M\, 2}\; \epsilon ^{T\, N}C^{-1}_{5}\tau ^{i}\epsilon ^{M}
\end{eqnarray*}

In order to make the WZ term to agree we have to set \( \alpha _{N}=\frac{1}{2\sqrt{2}}e^{i\pi /4}\theta _{N} \),
explicitly

\begin{equation}
\label{chang-ferm}
\widehat{\theta }=\frac{1}{2\sqrt{2}}e^{\frac{\iota \pi }{4}}\left( \begin{array}{c}
\left( \begin{array}{c}
\theta _{N}\\
-i\: \sigma _{2}\theta ^{*M}\, C_{5NM}
\end{array}\right) \otimes \epsilon ^{N}\\
0_{16}
\end{array}\right) 
\end{equation}

With this substitution our action (\ref{mia}) becomes

\begin{eqnarray}
S=-\frac{1}{2}\int d^{2}\xi \,  & \sqrt{-g}\, g^{\alpha \beta }\left[ \frac{y^{2}}{e^{2}}\left( \partial _{\alpha }x^{p}+2ie\: \widehat{\overline{\vartheta }}\Gamma ^{p}\partial _{\alpha }\widehat{\theta }\right) \left( \partial _{\beta }x_{p}+2ie\: \widehat{\overline{\vartheta }}\Gamma _{p}\partial _{\beta }\widehat{\theta }\right) +\frac{1}{e^{2}\, y^{2}}\partial _{\alpha }y^{t}\partial _{\beta }y_{t}\right]  & \nonumber \\
 & -4\frac{i}{e}y^{t}\: d\widehat{\overline{\vartheta }}\Gamma _{t}d\widehat{\theta } & \nonumber \label{loro} 
\end{eqnarray}
 which coincides with (\ref{loro}) exactly after performing an integration
by part of the WZ term, thus proving the exact equivalence of the two proposed
actions (at least on a word sheet without holes).  

\medskip{}
\textbf{\large Acknowledgments}{\large \par}

It is a pleasure to thank the Niels Bohr Institute for the hospitality.


\begin{thebibliography}{}
\bibitem{MatsaevTseytlin}R.R Metsaev and A.A. Tseytlin, Type IIB Superstring Action in \( AdS_{5}\times S_{5} \)
Background, hepth/9805028; \\
Supersymmetric D3 Brane Action \( AdS_{5}\times S_{5} \), hep-th/9506095
\bibitem{Kallosh}R. Kallosh, J. Rahmfeld and A. Rajaraman, Near Horizon Superspace, hep-th/9805217
\bibitem{Kall-fixing}R. Kallosh, Superconformal Actions in Killing Gauge, hep-th/9807206
\bibitem{Kall-act}R. Kallosh and J. Rahmfeld, The GS String Action on \( AdS_{5}\times S_{5} \),
hep-th/9808038
\bibitem{Kall-Tsey}R. Kallosh and A.K. Tseytlin, Simplifying Superstring Action on \( AdS_{5}\times S_{5} \),
hep-th/9808088
\bibitem{gruppoTo}G. Dall'Agata, D. Fabbri, C. Fraser, P. Fre', P. Termonia and M. Trigiante,
The \( Osp(8|4) \) Singleton Action from the Supermembrane, hep-th/9807115 
\bibitem{IP}I. Pesando, A \( \kappa  \) Gauge Fixed type IIB Superstring Action on \( AdS_{5}\times S_{5} \)
,hep-th/9802020
\end{thebibliography}
\end{document}